\pdfoutput=1
\documentclass[aps,prb,twocolumn,superscriptaddress]{revtex4}
\usepackage[latin1]{inputenc}
\usepackage{graphicx}
\usepackage{amsmath,amsfonts}

\begin{document}

\title{Domino plasmons for subwavelength terahertz circuitry}

\author{D. Martin-Cano}
\affiliation{Departamento de Fisica Teorica de la Materia
Condensada, Universidad Autonoma de Madrid, E-28049 Madrid, Spain}

\author{M. L. Nesterov}
\affiliation{Departamento de Fisica Teorica de la Materia
Condensada, Universidad Autonoma de Madrid, E-28049 Madrid, Spain}
\affiliation{A. Ya. Usikov Institute for Radiophysics and
Electronics, NAS of Ukraine, \\ 12 Academician Proskura Street,
61085 Kharkov, Ukraine}

\author{A. I. Fernandez-Dominguez}
\altaffiliation[Present address: ]{Physics Department, Blackett
Laboratory, Imperial College London, Prince Consort Road, London
SW7 2BZ, United Kingdom} \affiliation{Departamento de Fisica
Teorica de la Materia Condensada, Universidad Autonoma de Madrid,
E-28049 Madrid, Spain}

\author{F. J. Garcia-Vidal}
\affiliation{Departamento de Fisica Teorica de la Materia
Condensada, Universidad Autonoma de Madrid, E-28049 Madrid, Spain}

\author{L. Martin-Moreno}
\affiliation{Instituto de Ciencia de Materiales de Aragon (ICMA)
and Departamento de Fisica de la Materia Condensada,
CSIC-Universidad de Zaragoza, E-50009 Zaragoza, Spain}

\author{Esteban Moreno}
\email[Electronic address: ]{esteban.moreno@uam.es}
\affiliation{Departamento de Fisica Teorica de la Materia
Condensada, Universidad Autonoma de Madrid, E-28049 Madrid, Spain}

\begin{abstract}
A new approach for the spatial and temporal modulation of
electromagnetic fields at terahertz frequencies is presented. The
waveguiding elements are based on plasmonic and metamaterial
notions and consist of an easy-to-manufacture periodic chain of
metallic box-shaped elements protruding out of a metallic surface.
It is shown that the dispersion relation of the corresponding
electromagnetic modes is rather insensitive to the waveguide
width, preserving tight confinement and reasonable absorption loss
even when the waveguide transverse dimensions are well in the
subwavelength regime. This property enables the simple
implementation of key devices, such as tapers and power dividers.
Additionally, directional couplers, waveguide bends, and ring
resonators are characterized, demonstrating the flexibility of the
proposed concept and the prospects for terahertz applications
requiring high integration density.\end{abstract}

\maketitle

\section{Introduction}

Electromagnetic radiation in the terahertz (THz) regime is a
central resource for many scientific fields such as spectroscopy,
sensing, imaging, and communications. We are currently witnessing
the take off of THz
technologies~\cite{ferguson02,siegel02,tonouchi07} with
applications as diverse as astronomy~\cite{withington04},
medicine~\cite{siegel04}, or security~\cite{federici05}. Within
this general endeavour, the building of compact THz circuits would
stand out as an important accomplishment. This requires the design
of THz waveguides carrying tightly confined electromagnetic (EM)
modes, preferably with subwavelength transverse cross section.
Besides circuit integration and compact device design,
sub-$\lambda$ localization may be advantageous for waveguide THz
time-domain spectroscopy~\cite{zhang04} and
non-diffraction-limited imaging~\cite{cunningham08}. Although a
number of structures has been put forward, none of them passes all
the following requirements. First, structures should be easily
manufactured and, if possible, planar and monolithic. Second, as
above motivated, subwavelength transverse size is also needed.
Finally, absorption and bend losses should be small, and the
waveguides sufficiently versatile for the design of key functional
devices. Particularly important are in/out-couplers since they
work as the interface to external waves. In this context, compact
tapers able to laterally compress the modes down to the
sub-$\lambda$ level seem essential.

With these requirements in mind it is illuminating to compare
optical waveguides with those proposed for the THz range. Optical
fiber modes exhibit a cut-off diameter such that fibers cannot be
of subwavelength size and, moreover, modal size grows as this
cut-off is approached. In the THz regime dielectric
wires~\cite{jamison00} meet the same limitation. The photonic
crystal fiber approach~\cite{han02} shares the inability of
sub-$\lambda$ confinement, and low-index discontinuity
waveguides~\cite{nagel06} perform better regarding modal size but
are neither planar nor monolithic. Metals are an alternative when
operating at optical frequencies because surface plasmon
polaritons (SPPs) propagating on flat metal-dielectric interfaces
provide subwavelength confinement in the direction perpendicular
to the surface. This metallic route has been tried for the THz
regime, but the corresponding EM modes, known as Zenneck or
Sommerfeld waves, loose their confined character at these
frequencies~\cite{jeon06,wang04}.

Geometrically-induced surface plasmons~\cite{pendry04} supported
by periodically corrugated metallic surfaces overcome the low
localization of Zenneck waves. A number of waveguiding structures
based on this concept has been already
suggested~\cite{zhu08,maier06,fernandez09a,fernandez09b}, but they
have failed to meet simultaneously all the above mentioned
requirements. In this work we present structures based on this
metamaterial approach that, thank to their superior properties,
fulfill the requisite easy fabrication, subwavelength confinement,
low loss, and device flexibility.

\section{Results}

\begin{figure}[]
\includegraphics[scale=0.75]{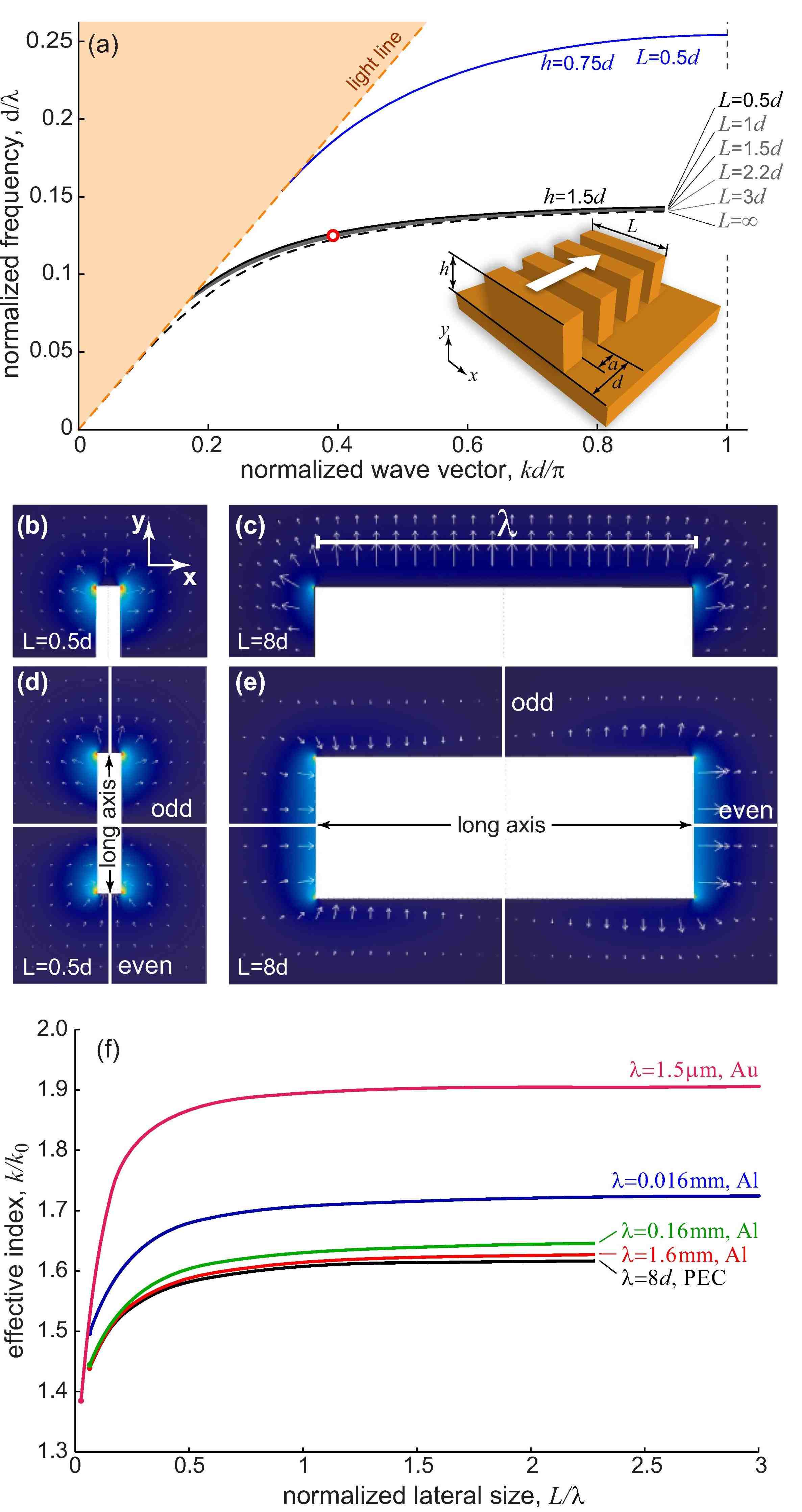}
\caption{\label{fig1} \textbf{Modal properties of domino
plasmons}. \textbf{a},~Dispersion relation of DPs for various
lateral widths $L$. Black and grey (blue) lines correspond to
height $h=1.5d$ ($h=0.75d$). Dashed line stands for infinitely
wide dominoes ($L=\infty$). Inset: diagram of the domino structure
and geometric parameters (the arrow depicts the mode propagation
direction). \textbf{b} and \textbf{c},~Modal shape of DPs:
transverse ($xy$) electric field (arrows) and horizontal ($xz$)
electric field (color shading) for DPs of height $h=1.5d$ and
widths $L=0.5d$ (\textbf{b}) and $L=8d$ (\textbf{c}). \textbf{d}
and \textbf{e},~Same as in panels \textbf{b} and \textbf{c}, but
now for a 1D array of free-standing metallic rods ($h=3d$). The
designations \emph{even} and \emph{odd} label the symmetries of
the modes with respect to the corresponding white lines. The
fields in \textbf{b}-\textbf{d} are plotted for $d/\lambda =
0.125$, marked with a red open dot in \textbf{a}, the white bar in
\textbf{c} being the wavelength (valid for panels
\textbf{b}-\textbf{d}). The field in \textbf{e} is computed for
the same $k$ as in panels \textbf{b}-\textbf{d}, corresponding now
to $d/\lambda = 0.056$. In panels \textbf{a}-\textbf{e} metals are
modelled as PECs. \textbf{f},~DP modal effective index as a
function of lateral dimension $L$ in units of wavelength. Various
operating frequency regimes are considered: $\lambda=1.6\,
\textrm{mm}$ (red), $\lambda=0.16\, \textrm{mm}$ (green),
$\lambda=0.016\, \textrm{mm}$ (blue), and $\lambda=1.5\, \mu
\textrm{m}$ (magenta). To compute panel \textbf{f}, a realistic
description of the metals is used. As described in the main text,
the periodicity $d$ is different for the various operating
frequencies, and $h=1.5d$, $a=0.5d$, $L=0.5d,\ldots
,24d$.}\end{figure}

The basic structure consists of a periodic arrangement of metallic
parallelepipeds standing on top of a metallic surface and
resembling a chain of domino pieces, see inset of
Fig.~\ref{fig1}(a). In contrast with corrugated
wires~\cite{maier06}, this is a planar and monolithic system and
should not pose significant manufacturing problems, its geometry
being much simpler than corrugated V-grooves~\cite{fernandez09a}
or wedges~\cite{fernandez09b}. The properties of its guided modes,
hereafter referred to as \emph{domino plasmons} (DPs), are mainly
controlled by the geometric parameters defining the dominoes,
\emph{i.e.}, periodicity $d$, parallelepiped height $h$, lateral
width $L$, and inter-domino spacing $a$.

First, the basic properties of DPs are described starting with the
dispersion relation. To gain physical insight we model the metal
as a perfect electric conductor (PEC). Later, realistic metals
will be considered, letting us to discern what is the frequency
regime where DPs display the most promising behaviour. The
numerical technique is described in the Methods section. Due to
the simple scaling properties of PECs, we choose the periodicity
$d$ as the length unit. The value of $a$ is not critical for the
properties of DPs and is set as $a=0.5d$ in this work. DP bands
present a typical plasmonic character, \emph{i.e.}, they approach
the light line for low frequencies and reach a horizontal
frequency limit at the edge ($k_{\textrm{edge}}=\pi /d$) of the
first Brillouin zone (Fig.~\ref{fig1}(a), notice that only
fundamental modes are plotted). In all cases DP modes lie outside
the light cone, a fact that explains their non-radiative
character. While the limit frequency of SPPs for large $k$ is
related to the plasma frequency, the corresponding value for DPs
is controlled by the geometry. The influence of the height $h$ is
clear: the band frequency rises for short dominoes ($h=0.75d$,
blue line) as compared with that of taller ones ($h=1.5d$, black
line). This can be understood as follows. When $L=\infty$,
dominoes become a one-dimensional (1D) array of grooves which
supports a geometrically-induced plasmon mode. Its dispersion
relation is represented, for $h=1.5d$, with a dashed line in
Fig.~\ref{fig1}(a). An approximation for the dispersion relation
in this limit, neglecting diffraction effects and for
$\lambda>>d$, is~\cite{garciavidal05}
\begin{equation}\label{eq1}
k=k_0\sqrt{1+\Bigl(\frac{a}{d}\Bigr)^2\tan^2(q_yh)},
\end{equation}
where $k$ is the modal wave vector, $k_0=2\pi/\lambda$, and the
$y$ component of the wave vector inside the grooves, $q_y$, is
related with the corresponding $x$ and $z$ components by
$q_x^2+q_y^2+q_z^2=k_0^2$. The EM fields inside the grooves are
independent of $x$ and $z$ for the chosen longitudinal
polarization and thus $q_x=q_z=0$, $q_y=k_0$. This means that
Eq.~\ref{eq1} presents an asymptote for $k_0 h=\pi/2$ or, in other
words, for a normalized frequency $d/4h$. Such dependence with $h$
explains the behaviour observed in Fig.~\ref{fig1}(a). For
$h=1.5d$, the value chosen in the rest of the paper, this
estimation for the limit normalized frequency provides a figure of
about $0.17$, not far from the numerically computed value of
$0.14$.

The most striking characteristic of DPs is their behaviour when
the lateral width $L$ is changed. All bands in the range
$L=0.5d,\ldots ,3d$ lie almost on top of each other
(Fig.~\ref{fig1}(a), grey curves). In other words, the modal
effective index, $n_{\textrm{eff}}=k/k_0$, is rather insensitive
to lateral width. If we now look at the numerically computed
dispersion relations of Fig.~\ref{fig1}(a) in the light of
Eq.~\ref{eq1}, it seems that $q_y\approx k_0$ for a wide range of
$L$. Remarkably, the bands remain almost unchanged even for
$L=0.5d$, whose modal size is well inside the subwavelength
regime, as observed in the field distribution plotted in panel (b)
for a narrow structure ($L=0.5d$). Notice that the white bar
representing $\lambda$ in panel (c) is valid for panels (b)-(d),
and the modal size, defined as the FWHM of the modulus of the
Poynting field distribution, is $0.21\, \lambda$ for panel (b).
The described behaviour is to be contrasted with that of
conventional plasmonic modes in the optical regime for which
sub-$\lambda$ lateral confinement is not a trivial
issue~\cite{verhagen09}. When the lateral size of the structure
supporting such modes becomes subwavelength, either the modal size
grows, or the effective index increases. This is best illustrated
for a free standing metallic stripe~\cite{berini00}. A stripe of
large lateral width essentially supports an SPP mode but, as this
width is decreased and becomes sub-$\lambda$, the plasmonic modes
that are not cut-off become either highly confined but quite
lossy, or otherwise highly delocalized.

The insensitivity of DPs to lateral size, which constitutes the
foundation of some of the devices presented later, can be linked
in an intuitive way to the structure geometry. The key role played
by the metallic substrate can be explained by comparison of DPs
and the modes of a Yagi-Uda structure, \emph{i.e.}, a periodic
array of free standing metallic
rods~\cite{sengupta59,krenn99,maier01}. For small $L$ the
dispersion of DPs of height $h$ is identical to that of the
fundamental mode of a Yagi-Uda structure of height $2h$, and the
field of the DP is simply the upper half of the fundamental mode
of a rod array of height $2h$, see Figs.~\ref{fig1}(b) and (d).
For large $L$ this correspondence does not hold anymore, as
demonstrated by comparison of panels (c) and (e) which display
completely different symmetries. For dominoes, a mode with the
same symmetry as the Yagi-Uda mode shown in panel (e) is forbidden
because the horizontal electric field has to be zero at the
metallic ground. This mode suppression is critical because the
frequency of the fundamental mode of a Yagi-Uda structure, which
is essentially controlled by the length of the long axis of the
rods, is very sensitive to the lateral width $L$, being very
different for short vertical rods and long horizontal rods. It is
similarly important that the air gaps between the dominoes are
laterally open. Let us imagine that, instead of being open, they
were laterally closed by PEC walls. Since the electric field
inside the gap points longitudinally from one domino to the next
one, it should then vanish at both PEC lateral walls, giving a
cosine-like $x$ dependence. It is then clear that the mode
frequency should increase when the lateral size $L$ is decreased,
arriving to the conclusion that the modes of such an structure
have $L$-sensitive bands. In contrast, for DPs the boundary
conditions at the laterally open gaps are more akin to perfect
magnetic walls, and the corresponding field is approximately
$x$-independent inside the air gaps even when $L\rightarrow 0$.

Let us scrutinize in more detail the role played by the lateral
dimension $L$, considering now realistic metals and paying
attention to the spectral regime. The periodicity $d$ is chosen to
set the operating wavelength in the desired region of the EM
spectrum, and $L$ is varied in the range $L=0.5d,\ldots ,24d$,
while the remaining parameters are kept constant ($a=0.5d$,
$h=1.5d$). Aluminum is selected for low frequencies, where metals
behave almost like PECs. In order to work at $\lambda=1.6\,
\textrm{mm}$, providing an operating angular frequency of the
order of 1 THz, we first consider $d=200\, \mu\textrm{m}$. The
evolution of the modal effective index as a function of the
lateral dimension normalized to the wavelength is plotted in
Fig.~\ref{fig1}(f). The curves corresponding to a PEC (black line)
and aluminum at $\lambda=1.6\, \textrm{mm}$ (red line) are, as
expected, almost identical. We can now quantify the sensitivity of
the effective index to $L$, its variation being only about 12\%
even when $L$ goes from $L=\infty$ to $L=0.5d=\lambda/16$,  well
inside the sub-$\lambda$ regime. The rapid variation of
$n_{\textrm{eff}}$ when $L<<d$ can be understood with the help of
Eq.~\ref{eq1}. In this limit the EM fields start to be
$z$-dependent inside the air gaps, as verified by inspection of
the fields (not shown here), letting us to conclude that $q_z\neq
0$ and consequently $q_y<k_0$. Then, if we assume that
Eq.~\ref{eq1} is still valid in this limit, it predicts that the
modal effective index is reduced when $L\rightarrow0$. To
investigate the performance of DPs at higher frequencies, the
structures have been scaled down by factors 1/10 and 1/100. The
fact that the curves corresponding to $\lambda=0.16\, \textrm{mm}$
(green line) and $\lambda=0.016\, \textrm{mm}$ (blue line) do not
lie on top of the previous ones is a signature of the departure of
aluminum from the PEC behaviour. Nevertheless, even at
$\lambda=0.016\, \textrm{mm}$, the variation of the effective
index is still smaller than 15\%, a property that can be exploited
for taper design in the far infrared, albeit with slightly reduced
performance as compared with the THz case. When the operating
frequency moves to the telecom regime ($\lambda=1.5\, \mu
\textrm{m}$, magenta line) the variation of the effective index is
much larger (about 38\%). Let us mention in passing that, for this
wavelength, we have used gold because it has lower loss than
aluminum, and that the geometric scaling had to be different (the
periodicity is now $d=0.13\, \mu \textrm{m}$) because with the
same simple scaling as before no DP mode is supported due to the
value of the permittivity of gold. The important message of panel
(f) is that, although a variation of $n_{\textrm{eff}}$ begins to
be noticeable when the lateral dimension $L$ goes below $\lambda
/2$, DP bands are fairly insensitive to $L$ in the range
$L=0.5d,\ldots ,24d$ when operating at low frequencies. Such
property has, to our knowledge, never been reported for
geometrically-induced plasmon modes in corrugated wedges,
channels, or wires.

\begin{figure}[]
\includegraphics[scale=0.84]{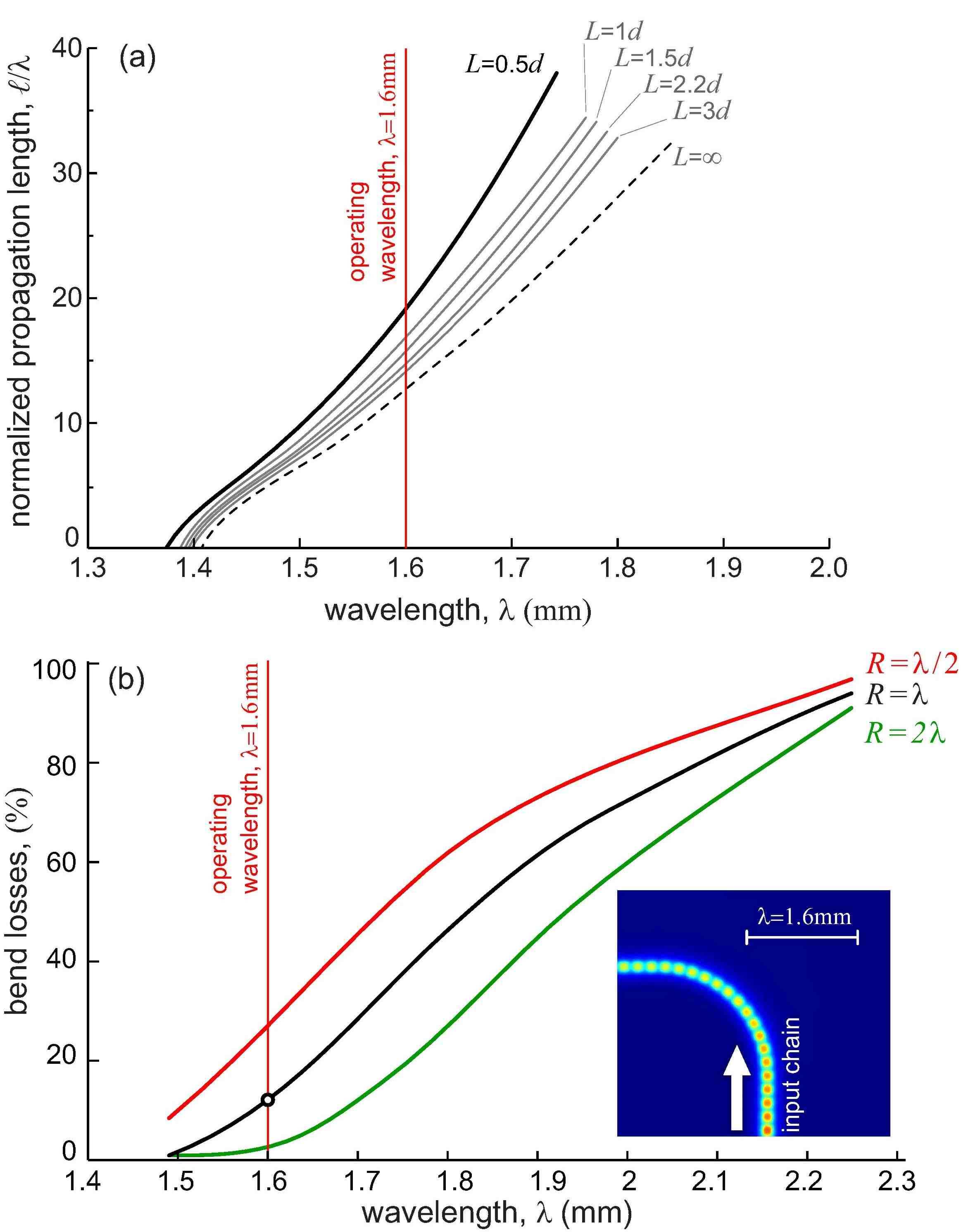}
\caption{\label{fig2} \textbf{Absorption (ohmic) and bend
(radiation) losses}. \textbf{a},~Normalized propagation length of
DPs in rectilinear guides of various $L$ as a function of
$\lambda$ ($h=1.5d$, $a=0.5d$, $d=200\, \mu\textrm{m}$).
\textbf{b},~Bend loss of DPs for three radii of curvature as a
function of $\lambda$ ($h=1.5d$, $a=0.5d$, $L=0.5d$, $d=200\,
\mu\textrm{m}$). Inset:~Poynting vector field (modulus)
distribution in a horizontal plane slightly above ($30\,
\mu\textrm{m}$) the height of the bend (top view). The chosen
wavelength and radius of curvature are marked with an open black
dot in panel (b). The red solid vertical lines in both panels
indicate the operating wavelength used later, $\lambda=1.6\,
\textrm{mm}$.}\end{figure}

Now that we have identified the THz range as the most promising
operation regime, we will characterize the corresponding
absorption (ohmic) and bending losses. Figure~\ref{fig2}(a)
displays the propagation length ($\ell=[2 \textrm{Im}(k)]^{-1}$)
of DPs as a function of wavelength for various widths $L$
($d=200\, \mu \textrm{m}$ for all curves). The only source of
damping in the domino structure is absorption by the metal as no
radiation losses occur in such straight waveguide. Propagation
length is very short for small $\lambda$, when the Brillouin zone
edge is reached, but it rises for increasing wavelength as the
dispersion curve approaches the light line. Several lateral widths
$L$ are considered and it is observed that DPs in structures of
smaller $L$ travel longer distances before being damped. The
radiation losses occurring in a waveguide of $L=0.5d$ with a 90
degree bend are analyzed in panel (b). In this case the metal is
modelled as a PEC for easier evaluation of radiation losses. Data
for three radii of curvature of the bend, $R=\lambda/2, \lambda,
2\lambda$, are presented. In all three cases, losses are low for
small $\lambda$ but they grow as the wavelength is increased and
the modal wave vector approaches $k_0$. The dependence with the
radius of curvature of the bend is not surprising either,
radiation losses being higher for smaller radius of curvature.
Consideration of absorption and bending losses suggests that
$\lambda=1.6\, \textrm{mm}$ can be an appropriate operating
wavelength. For $\lambda=1.6\, \textrm{mm}$ and $L=0.5d=100\,
\mu\textrm{m}$, the modal size is $0.34\, \textrm{mm}$,
\emph{i.e.}, about $\lambda/5$. The propagation length is about
$20\lambda$, much larger than the size of the devices presented
later, and the bend loss for $R=\lambda$ is reasonably low, about
10\%. Notice that for corrugated V-grooves, the bend loss for
$R=2\lambda$ is around 50\%~\cite{fernandez09a}.

\begin{figure}[]
\includegraphics[scale=0.92]{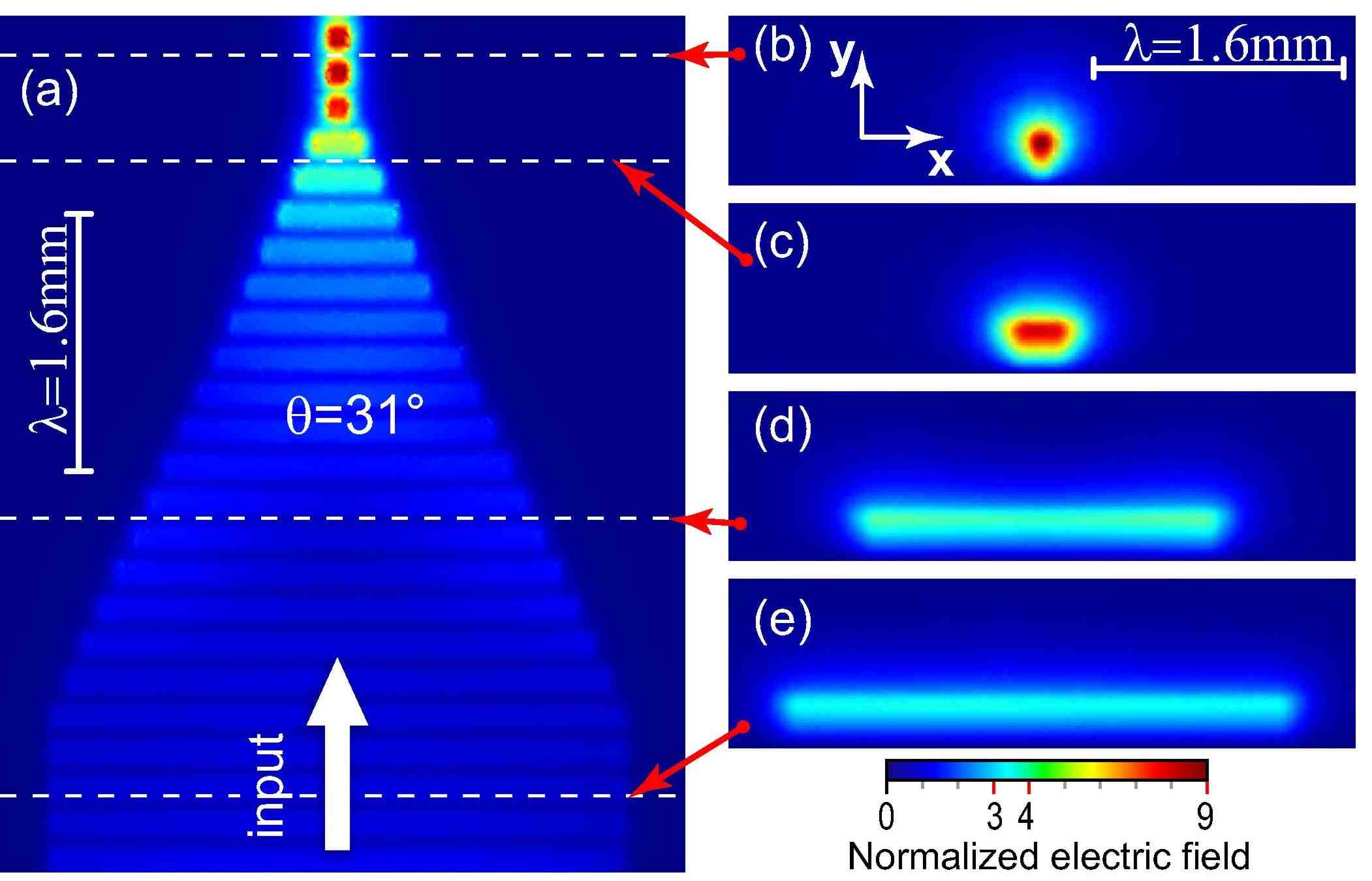}
\caption{\label{fig3} \textbf{Subwavelength concentration of a
domino plasmon}. \textbf{a},~Poynting vector field (modulus)
distribution in a horizontal plane slightly above ($30\,
\mu\textrm{m}$) the height of the tapered domino structure (top
view). The lateral width is tapered from $L_{\textrm{in}}=16d$ to
$L_{\textrm{out}}=0.5d$ ($h=1.5d$, $a=0.5d$, $d=200\,
\mu\textrm{m}$, $\lambda=1.6\, \textrm{mm}$).
\textbf{b}-\textbf{e},~Amplitude of electric field in transverse
vertical planes (longitudinal views) at the locations shown by
white dashed lines in \textbf{a}. The white bar in \textbf{b}
showing the operating wavelength is valid for the last four
panels.}\end{figure}

In the rest of the paper we present a number of devices enabling
the spatial and temporal modulation of EM fields. The operating
wavelength is $\lambda=1.6\, \textrm{mm}$ and $d=200\, \mu
\textrm{m}$. In all ensuing computations the metal is considered
as a PEC, to easily quantify the device radiation loss. Waveguide
tapering~\cite{maier06,rusina08,liang08,fernandez09b} is
interesting for field concentration and amplification, both
important requirements for imaging
applications~\cite{kawata09,ishihara06} and circuit
integration~\cite{ishikawa09}. Tapering is expected to be easy in
domino structures due to the insensitivity to lateral width $L$ of
the dispersion relation of DPs. This hypothesis is confirmed in
Fig.~\ref{fig3}(a), which shows a top view of the propagation of
power along a tapered domino structure. The lateral widths of the
waveguide are $L_{\textrm{in}}=16d$ at the input port (bottom) and
$L_{\textrm{out}}=0.5d$ at the output port (top). The length of
the taper is $16d$, \emph{i.e.}, two wavelengths at the operating
frequency, corresponding to a taper semi-angle of
$\theta=31^{\circ}$. Remarkably, reflection is smaller than 2\%
and only 5\% of the incoming power is lost as radiation due to the
shrinkage of the lateral dimension along the taper. Panels (b)-(e)
are cross sections at different positions along the taper, vividly
showing the process of field subwavelength concentration and
enhancement. As mentioned above, the same design does not work in
the telecom regime due to the sensitivity of DP bands to the
lateral dimension $L$ at those frequencies. A more adiabatic
taper, \emph{i.e.}, with smaller change of $n_{\textrm{eff}}$ per
unit length, would not circumvent the problem because absorption
loss is high in the telecom regime (for instance, for
$\lambda=1.5\, \mu \textrm{m}$, $d=0.13\, \mu \textrm{m}$, and
$L=16d$, the propagation length is $\ell=1.2\, \mu \textrm{m}$).

\begin{figure}[]
\includegraphics[scale=0.81]{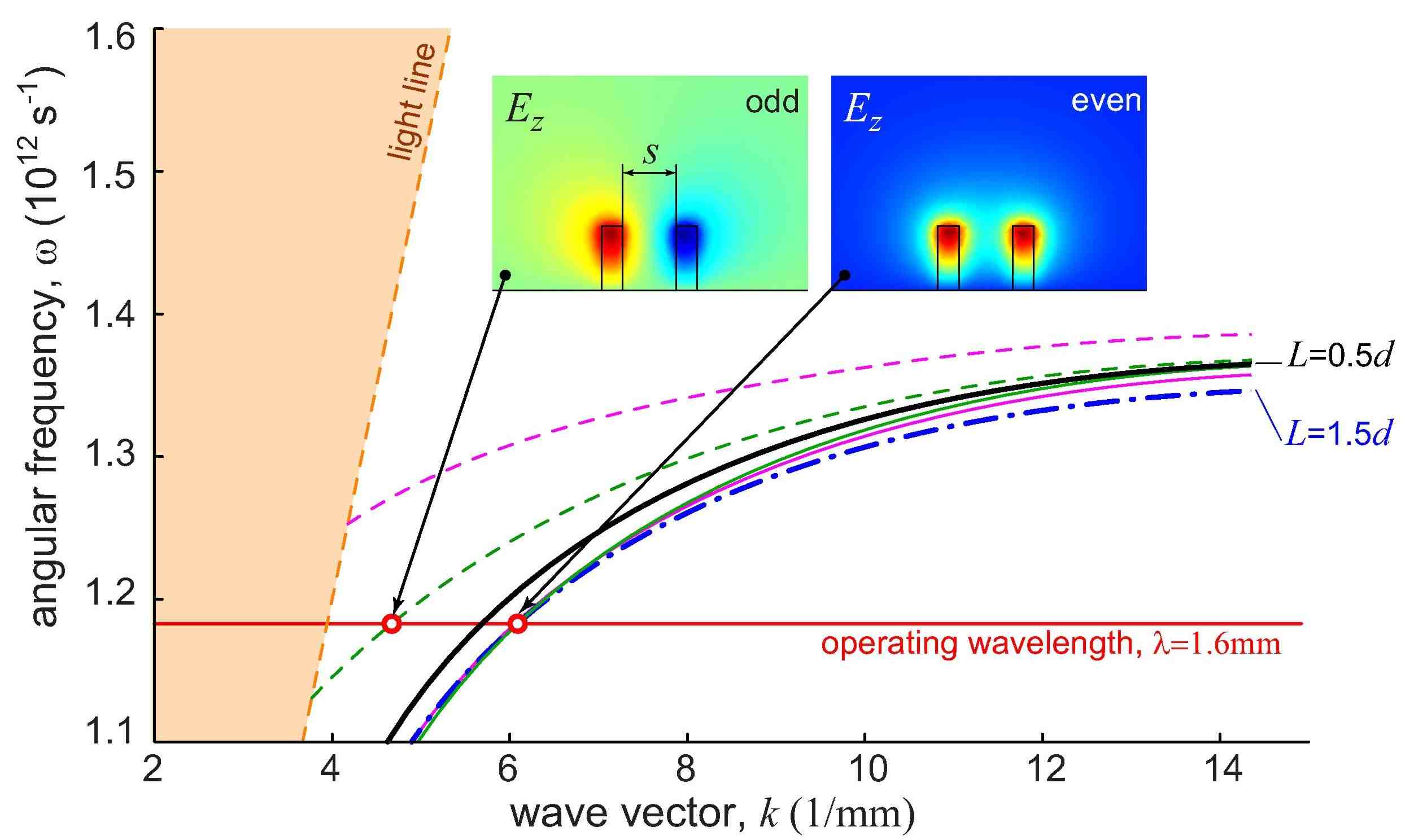}
\caption{\label{fig4} \textbf{Dispersion relations corresponding
to one and two parallel domino structures}. Black solid (blue
dashed-dotted) line is for DP of $L=0.5d$ ($L=1.5d$). Magenta
solid (dashed) line is for the even (odd) supermode of parallel
domino structures separated a distance $s=0.5d$, whereas green
solid (dashed) line is the corresponding supermode for $s=1.25d$.
The individual dominoes in double structures have $L=0.5d$. The
red solid horizontal line indicates the operating wavelength,
$\lambda=1.6\, \textrm{mm}$, used later. Insets: longitudinal
electric field for the odd and even supermodes of two parallel
domino structures separated a distance $s=1.25d$ at the operating
wavelength, and displayed in a transverse cross section lying at
the center of the inter-domino gaps. For all structures $h=1.5d$,
$a=0.5d$, $d=200\, \mu\textrm{m}$.}\end{figure}

\begin{figure*}[]
\includegraphics[scale=1.4]{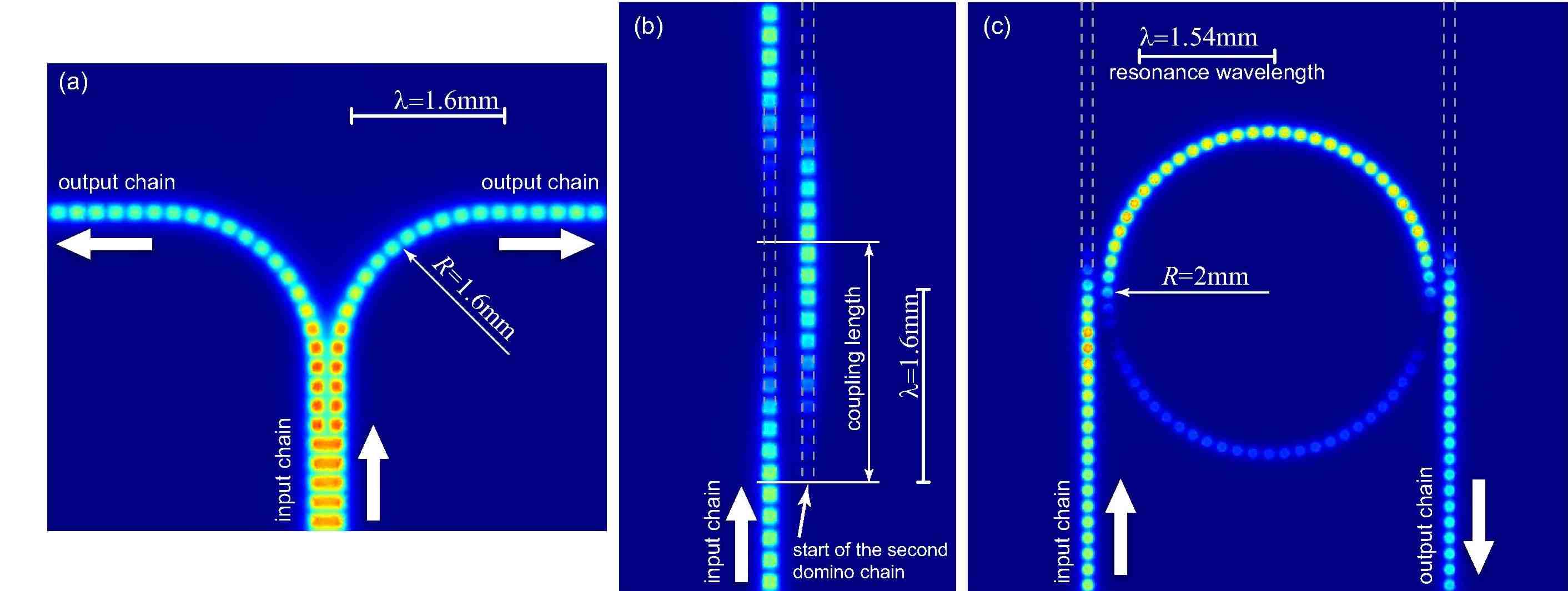}
\caption{\label{fig5} \textbf{Domino plasmon devices}. Top view
of: \textbf{a},~Power divider. \textbf{b},~Directional coupler.
\textbf{c},~Waveguide ring resonator. The Poynting vector
(modulus) is displayed in a horizontal plane slightly above (about
$30\, \mu\textrm{m}$) the height of the domino structures. White
bars show the operating wavelength in each panel. The various
geometric parameters are described in the main text, the
periodicity being $d=200\, \mu\textrm{m}$ in all
cases.}\end{figure*}

We now discuss how to construct power dividers and directional
couplers~\cite{huang08,nesterov08}. Figure~\ref{fig4} plots the
dispersion relation of the modes supported by one and two parallel
domino chains. Black solid line corresponds to the DP of $L=0.5d$.
For two parallel waveguides (both with $L=0.5d$), interaction
gives rise to symmetric and antisymmetric supermodes which, for a
separation $s=0.5d$, are shown by magenta curves with the solid
line depicting the even supermode whereas the dashed one
represents the odd supermode. Blue dashed-dotted line corresponds
to a wider structure of $L=1.5d$ and, importantly, this band stays
very close to that of the $s=0.5d$ even supermode. Due to this
good overlapping, a simple linking of a wide domino structure with
$L=1.5d$ to two parallel domino chains (widths $L=0.5d$ and
separation $s=0.5d$) should not result in an important impedance
mismatch. This is verified in Fig.~\ref{fig5}(a), displaying a top
view of a power divider based on the previous idea. The input port
($L=1.5d$) at the bottom feeds the wide domino which, after 5
periods, is linked to two parallel domino structures ($L=0.5d$,
$s=0.5d$). Due to symmetry only the even supermode is excited. The
lateral separation $s$ between the individual waveguides is then
increased, followed by $90^{\circ}$ waveguide bends. Simulations
show that reflection is also negligible in this power splitter.
Total radiation loss in the device is equal to that happening in
the bends (previously discussed in Fig.~\ref{fig2}(b)), each being
5\% of the incoming power, which is small taking into account that
the bend radius is equal to the operating wavelength.

Coupling between parallel domino structures can also be easily
understood with the help of Fig.~\ref{fig4}, displaying with green
lines the even (solid) and odd (dashed) supermodes for parallel
domino chains separated a distance $s=1.25d$. The insets show the
modal shape of both modes. A larger $s$ is chosen in this
simulation to prevent the cut-off of the odd supermode. The
coupling length needed to exchange the carried power between the
waveguides is estimated as
$\pi/|k_{\textrm{even}}-k_{\textrm{odd}}|=2.05\, \textrm{mm}$.
This value agrees with the result found in the simulation of a
directional coupler. Figure~\ref{fig5}(b) displays a top view of
two parallel domino chains ($L=0.5d$) separated a distance
$s=1.25d$. A mode is launched in the left bottom waveguide and
characteristic beatings with the predicted coupling length are
observed when a second waveguide is located close to it.

As a final example a waveguide ring resonator is demonstrated,
able to route various frequencies to different output
ports~\cite{bozhevolnyi06}. To show once more the robustness of
our proposal, this device has been designed with posts of circular
cross section as seen from above (radius $r=0.28d$), instead of
square cross section parallelepipeds. We have checked that similar
results are obtained for the square geometry. The top view of this
device is displayed in Fig.~\ref{fig5}(c). For the chosen
geometric parameters (ring radius $R=2\, \textrm{mm}$, minimum
separation in the coupling sections $s=1.25d$), it is possible to
drop a resonant wavelength $\lambda=1.54\, \textrm{mm}$ at the
output port. Non-negligible radiation losses only occur at the
bends in the waveguide ring and they are about 25\% of the input
power, reasonably small considering that no optimization was
attempted.

In summary, we have presented a new class of surface
electromagnetic modes, termed domino plasmons, which feature an
extraordinary characteristic: their insensibility to the lateral
dimension of the structure supporting them. We have shown how the
guiding properties of these modes enable the planar routing of THz
radiation at a deep subwavelength level. Finally, the flexibility
and versatility of waveguides based on domino plasmons have been
demonstrated through the implementation of a variety of functional
devices, which may thus prove very useful as a new concept for
subwavelength THz circuitry.

\section{Methods}

All results in this paper have been obtained by means of numerical
simulations performed with the Finite Element Method (FEM) using
commercial (COMSOL Multiphysics) software. A summary of the main
modelling details follows. The technique is a volume FEM operating
in frequency domain. Due to the various length scales involved,
highly non-uniform meshes are used. In order to ensure an adequate
representation of the electromagnetic fields, the size of the
elements is a fraction of the skin depth inside the metallic
parallelepipeds, a fraction of the characteristic geometric
dimension (\emph{e.g.}, $a$) in the neighborhood of the structure,
and a fraction of the operating wavelength at the boundaries of
the simulation domain. The mesh size is refined until the results
are stable with an accuracy of 1\%. For instance, in the case of
band structure computations, the convergence criterium is based on
the value of both the real and imaginary parts of the wave vector.
The final mesh is different for each simulation, but typical
values of the tetrahedra sizes in different regions of the
simulation domain are of the order of 1/5 of the skin depth and
characteristic geometric dimension, and 1/10 of the wavelength.
The typical number of degrees of freedom lies between
$3\times10^5$ and $3\times10^6$, a range in which the matrix
equations are inverted with direct solvers. For the computation of
band structures, the corresponding eigenvalue problem is posed in
a single unit cell where Bloch boundary conditions together with
Scattering boundary conditions are used. For device modelling, the
excitation of the corresponding scattering problem is provided by
Port boundary conditions, the ports being fed by the solutions of
the corresponding eigenvalue problem. Open space is mimicked with
Perfectly Matched Layers or/and Scattering boundary conditions.
For the simulation of ideal metals, Perfect Electric Conductor
(PEC) boundary conditions are employed. For realistic metals the
dielectric permittivities are taken from
Ref.~[\onlinecite{ordal83}] for Al
($\epsilon_{\textrm{Al}}=-3.39\times10^4+i3.5\times10^6$ at
$\lambda=1.6\, \textrm{mm}$), and Ref.~[\onlinecite{johnson72}]
for Au ($\epsilon_{\textrm{Au}}=-103+i8.7$ at $\lambda=1.5\, \mu
\textrm{m}$). At low frequencies it was often sufficient to
simulate the domains outside the conductors together with Surface
Impedance Boundary Conditions, as verified by comparison with
simulations fully accounting for the EM fields inside the metals.

\end{document}